\documentclass[5p,times,sort&compress]{elsarticle}

\journal{Journal of Quantitative Spectroscopy \& Radiative Transfer}

\usepackage{graphicx}
\usepackage{xcolor}
\usepackage[colorlinks=true,linkcolor=blue,urlcolor=blue,citecolor=blue,bookmarks=true,bookmarksopenlevel=2]{hyperref}
\usepackage{amsmath}
\usepackage[utf8]{inputenc}
\usepackage[T1]{fontenc}
\usepackage{microtype}
\usepackage{bm}
\usepackage{array,tabularx,booktabs}
\usepackage{rotating}

\newcolumntype{L}{>{\raggedright\arraybackslash}X}
\newcolumntype{R}{>{\raggedleft\arraybackslash}X}
\newcolumntype{C}{>{\centering\arraybackslash}X}

\newcommand{\ket}[1]{\left\vert#1\right\rangle}

\usepackage{siunitx}
\usepackage{xspace}
\newcommand{\WCu}{W\textsuperscript{45+}\xspace}
\newcommand{\WNi}{W\textsuperscript{46+}\xspace}
\newcommand{\WCl}{W\textsuperscript{57+}\xspace}
\newcommand{\Wviii}{W\textsuperscript{7+}\xspace}
\newcommand{\grasp}{\textsc{Grasp2k}\xspace}
\newcommand{\mcdfgme}{\textsc{Mcdfgme}\xspace}
\newcommand{\fac}{\textsc{Fac}\xspace}
\newcommand{\qedmod}{\textsc{Qedmod}\xspace}

\begin{document}

\begin{frontmatter}
\title{Breit and QED contributions in atomic structure calculations of tungsten ions}
\author{Karol Kozio{\l}} 
\ead{Karol.Koziol@ncbj.gov.pl}
\address{Narodowe Centrum Bada\'{n} J\k{a}drowych (NCBJ), Andrzeja So{\l}tana 7, 05-400 Otwock-\'{S}wierk, Poland}

\begin{abstract}
The \fac, \grasp, and \mcdfgme codes are compared in three case studies of the radiative transitions occurring in tungsten ions: (i) Ni1 and Ni2 lines in Ni-like tungsten, (ii) $3p_{3/2}{-}3p_{1/2}$ fine splitting in Cl-like tungsten, and (iii) $K\alpha_{1}$ and $K\alpha_{2}$ lines in W VIII. 
Various approaches to including the Breit interaction term and QED corrections in atomic calculations are examined. 
Electron correlation effects are also investigated and compared to the Breit and QED contributions. 
The data presented here may be used to estimate theoretical uncertainties relevant to interpretation of high-resolution spectroscopic data. 
\end{abstract}


\end{frontmatter}

\section{Introduction}

The investigation of tungsten ions is of great importance in theoretical and applied atomic physics. 
Firstly, high-$Z$ atoms such as tungsten, are used to probe relativistic and quantum-electrodynamics (QED) effects \cite{Gillaspy2010} and have been suggested as potential candidates for testing the time variation of the fine structure constant \cite{Berengut2011}. 
Secondly, tungsten is chosen as a plasma facing material in modern large tokamaks, such as JET (Joint European Torus) and ITER (International Thermonuclear Experimental Reactor), and thus constitutes the majority of the impurity ions in the tokamak plasma. 
Therefore, spectroscopic studies of tungsten ions provide a diagnostic tool relevant to a wide range of electron temperatures \cite{Putterich2008,Ralchenko2006}. 
High-precision atomic calculations are required to interpretation of complex spectra of high-charged tungsten ions and they constitute the solid base for the farther benchmark modelling of the radiation emission from tokamak plasmas. 
Several codes are used to predict the atomic structure and transition probabilities of ions that are of interest in plasma research, such as \textsc{Relac} \cite{Klapisch1977}, the Cowan code \cite{Cowan1981}, \textsc{Hullac} \cite{Bar-Shalom2001}, \textsc{Grasp} \cite{Dyall1989} and \grasp \cite{Jonsson2007,Jonsson2013}, \mcdfgme \cite{Gorceix1987,Indelicato1987}, \textsc{Rmbpt} \cite{Safronova1996}, and \fac \cite{Gu2008}. Recently, two of these, \fac and \grasp, have become the most widely used codes. When comparing results obtained from \fac and \grasp calculations, it is crucial to know which theoretical contributions are included in the calculations and what may potentially produce discrepancies. Hence, the aim of this article is to discuss the differences in theoretical contributions taken into account by the \fac, \grasp, and \mcdfgme codes. 
The relative roles of the electron correlation contribution and the Breit and QED contributions is also examined.

Both the \grasp and the \mcdfgme codes use the Multi-Configuration Dirac-Hartree-Fock (MCDHF) approach. The methodology of MCDHF calculations has been presented in many papers; see, e.g.,  \cite{Grant2007}. 
The \textsc{Grasp} (General-Purpose Relativistic Atomic Structure Program) code was developed by the Grant group at University of Oxford and recently improved by Froese Fischer, J{\"o}nsson, and collaborators in order to perform large-scale Configuration Interaction (CI) calculations.  
The \mcdfgme (Multi Configuration Dirac Fock and General Matrix Element) code was developed by Desclaux and Indelicato in France, and takes into account the Breit and QED corrections in a detailed way.  
The \fac (Flexible Atomic Code), utilising the modified multiconfigurational Dirac-Hartree-Fock-Slater (DHFS) method, was developed by M.\ F.\ Gu at Stanford University for speed, multi-utility, and collisional-radiative modelling. The main difference between the DHF and DHFS methods is that DHFS approximates the non-local DHF exchange potential by a local potential. Since the DHFS method uses an approximate form of the electron-electron interaction potential, it is commonly considered to be less accurate than the more sophisticated MCDHF method. This assumption is examined in the present work.  
The main aim of the research presented here is to estimate the theoretical uncertainties relevant to the interpretation of high-resolution spectroscopic data. 

\begin{sidewaystable*}
\setlength{\tabcolsep}{4pt}
\caption{\label{tab:ni12}Various theoretical contributions to the energy of Ni1 and Ni2 transitions in Ni-like tungsten ion (eV).}
\small
\begin{tabular*}{\linewidth}{@{\extracolsep{\fill}}l *{12}{>{\small\arraybackslash}r}}
\toprule
& \multicolumn{4}{c}{$\ket{[Mg]3p^63d^{10}}_{J=0}$ ($^1S_{0}$)} & \multicolumn{4}{c}{$\ket{[Mg]3p_{1/2}^2 3p_{3/2}^3 3d^{10} 4d_{3/2}^1}_{J=1}$ ($^3P_{1}$)} & \multicolumn{4}{c}{$\ket{[Mg]3p_{1/2}^2 3p_{3/2}^3 3d^{10} 4d_{5/2}^1}_{J=1}$ ($^1P_{1}$)} \\
\cmidrule{2-5}\cmidrule{6-9}\cmidrule{10-13}
 &   \multicolumn{1}{c}{\grasp}  &  \multicolumn{1}{c}{\grasp}  &   \multicolumn{1}{c}{\mcdfgme}  &   \multicolumn{1}{c}{\fac}  &   \multicolumn{1}{c}{\grasp}  &  \multicolumn{1}{c}{\grasp}  &   \multicolumn{1}{c}{\mcdfgme}  &   \multicolumn{1}{c}{\fac}  &   \multicolumn{1}{c}{\grasp}  &  \multicolumn{1}{c}{\grasp}  &   \multicolumn{1}{c}{\mcdfgme}  &   \multicolumn{1}{c}{\fac} \\
Contribution  &    &  \multicolumn{1}{c}{+\qedmod}  &    &    &    &  \multicolumn{1}{c}{+\qedmod}  &    &    &    &  \multicolumn{1}{c}{+\qedmod}  &    &   \\
\midrule
Dirac-Fock  & -399290.21 &    & -399287.25 & -399266.93 & -396923.31 &    & -396920.38 & -396900.97 & -396898.12 &    & -396895.20 & -396875.68\\
Breit($\omega=0$)+Rec.  & 444.25 &    & 443.17 & 444.66 & 440.02 &    & 438.89 & 440.73 & 439.30 &    & 438.20 & 440.02\\
\hspace{1em}Mag.  &    &    & 493.73 &    &    &    & 488.60 &    &    &    & 487.82 &   \\
\hspace{1em}Ret.($\omega=0$)  &    &    & -50.66 &    &    &    & -49.81 &    &    &    & -49.73 &   \\
\hspace{1em}Recoil  &    &    & 0.10 &    &    &    & 0.10 &    &    &    & 0.10 &   \\
Breit($\omega>0$)  & -8.84 &    & -8.80 &    & -8.44 &    & -8.40 &    & -8.45 &    & -8.40 &   \\
VP  &    &    &    &    &    &    &    &    &    &    &    &   \\
\hspace{1em}VP11  &    & -71.10 & -70.84 & -71.76 &    & -71.12 & -70.86 & -71.79 &    & -71.12 & -70.86 & -71.79\\
\hspace{1em}VP11+21  & -71.66 &    & -71.39 &    & -71.69 &    & -71.41 &    & -71.69 &    & -71.41 &   \\
\hspace{1em}VP11+21+13  &    & -68.80 & -68.79 &    &    & -68.83 & -68.81 &    &    & -68.83 & -68.81 &   \\
SE  &    &    &    &    &    &    &    &    &    &    &    &  \\
\hspace{1em}Welt.  & 374.64 &    & 374.60 & 372.94 & 374.22 &    & 374.10 & 372.54 & 374.26 &    & 374.12 & 372.55\\
\hspace{1em}dens.  & 380.15 &    &    &    & 379.73 &    &    &    & 379.76 &    &    &   \\
\hspace{1em}mod.  &    & 374.91 &    &    &    & 374.46 &    &    &    & 374.51 &    &   \\
QED h.o.  &    &    & -1.15 &    &    &    & -1.15 &    &    &    & -1.15 &   \\
Total(Welt.)  & -398551.82 &    & -398548.20 & -398521.09 & -396189.20 &    & -396185.75 & -396159.49 & -396164.70 &    & -396161.24 & -396134.90\\
Total(dens.)  & -398546.31 &    &    &    & -396183.69 &    &    &    & -396159.20 &    &    &   \\
Total(mod.)  &    & -398548.69 &    &    &    & -396186.11 &    &    &    & -396161.59 &    &   \\
\midrule
 &  \multicolumn{4}{c}{Ni1 (${}^1P_1 \to {}^1S_0$)}  &  \multicolumn{4}{c}{Ni2 (${}^3P_1 \to {}^1S_0$)}   &    &    &    &     \\
\cmidrule{2-5}\cmidrule{6-9}
 &   \multicolumn{1}{c}{\grasp}  &  \multicolumn{1}{c}{\grasp}  &   \multicolumn{1}{c}{\mcdfgme}  &   \multicolumn{1}{c}{\fac}  &   \multicolumn{1}{c}{\grasp}  &  \multicolumn{1}{c}{\grasp}  &   \multicolumn{1}{c}{\mcdfgme}  &   \multicolumn{1}{c}{\fac} \\
Contribution  &    &  \multicolumn{1}{c}{+\qedmod}  &    &    &    &  \multicolumn{1}{c}{+\qedmod}  &    &  &&&& \\
\midrule
Dirac-Fock   & 2392.09 &      & 2392.05 & 2391.25 & 2366.90 &      & 2366.87 & 2365.96 &&&& \\
Breit($\omega=0$)+Rec.   & -4.95 &      & -4.97 & -4.64 & -4.23 &      & -4.28 & -3.93 & &&& \\
\hspace{1em}Mag.   &      &      & -5.91 &      &      &      & -5.14 &      & &&& \\
\hspace{1em}Ret.($\omega=0$)   &      &      & 0.93 &      &      &      & 0.85 &      & &&& \\
\hspace{1em}Recoil   &      &      & 0.00 &      &      &      & 0.00 &      & &&& \\
Breit($\omega>0$)   & 0.38 &      & 0.39 &      & 0.40 &      & 0.40 &      & &&& \\
VP   &      &      &      &      &      &      &      &      & &&& \\
\hspace{1em}VP11   &      & -0.02 & -0.02 & -0.03 &      & -0.02 & -0.02 & -0.03 & &&& \\
\hspace{1em}VP11+21   & -0.02 &      & -0.02 &      & -0.02 &      & -0.02 &      & &&& \\
\hspace{1em}VP11+21+13   &      & -0.02 & -0.02 &      &      & -0.02 & -0.02 &      & &&& \\
SE   &      &      &      &      &      &      &      &      & &&& \\
\hspace{1em}Welt.   & -0.38 &      & -0.48 & -0.39 & -0.41 &      & -0.51 & -0.40 & &&& \\
\hspace{1em}dens.   & -0.39 &      &      &      & -0.42 &      &      &      & &&& \\
\hspace{1em}mod.   &      & -0.40 &      &      &      & -0.46 &      &      & &&& \\
QED h.o.   &      &      & 0.00 &      &      &      & 0.00 &      & &&& \\
Total(Welt.)   & 2387.12 &      & 2386.96 & 2386.19 & 2362.62 &      & 2362.45 & 2361.61 & &&& \\
Total(dens.)   & 2387.11 &      &      &      & 2362.62 &      &      &      & &&& \\
Total(mod.)   &      & 2387.10 &      &      &      & 2362.58 &      &      & &&& \\
\bottomrule
\end{tabular*}
\end{sidewaystable*}

\begin{sidewaystable}
\setlength{\tabcolsep}{4pt}
\caption{\label{tab:cl3p}Various theoretical contributions to the energy of $3p_{3/2}-3p_{1/2}$ fine splitting in Cl-like tungsten ion (eV).}
\small
\begin{tabular*}{\linewidth}{@{\extracolsep{\fill}}l *{12}{>{\small\arraybackslash}r}}
\toprule
& \multicolumn{4}{c}{$3p_{1/2}$} & \multicolumn{4}{c}{$3p_{3/2}$} & \multicolumn{4}{c}{$3p_{3/2}-3p_{1/2}$} \\
\cmidrule{2-5}\cmidrule{6-9}\cmidrule{10-13}
  &    \multicolumn{1}{c}{\grasp}   &   \multicolumn{1}{c}{\grasp}   &    \multicolumn{1}{c}{\mcdfgme}   &    \multicolumn{1}{c}{\fac}   &    \multicolumn{1}{c}{\grasp}   &   \multicolumn{1}{c}{\grasp}   &    \multicolumn{1}{c}{\mcdfgme}   &    \multicolumn{1}{c}{\fac}   &    \multicolumn{1}{c}{\grasp}   &   \multicolumn{1}{c}{\grasp}   &    \multicolumn{1}{c}{\mcdfgme}   &    \multicolumn{1}{c}{\fac} \\
Contribution   &      &   \multicolumn{1}{c}{+\qedmod}   &      &      &      &   \multicolumn{1}{c}{+\qedmod}   &      &      &      &   \multicolumn{1}{c}{+\qedmod}   &      &    \\
\midrule &  &  &  &  &  &  &  &  &  &  &  & \\
Dirac-Fock   & -346399.48 &      & -346396.71 & -346379.23 & -346752.33 &      & -346749.55 & -346732.10 & 352.85 &      & 352.84 & 352.87\\
Breit($\omega=0$)+Rec.   & 400.90 &      & 399.90 & 401.31 & 404.47 &      & 403.45 & 405.12 & -3.58 &      & -3.56 & -3.81\\
\hspace{1em}Mag.   &      &      & 442.39 &      &      &      & 445.92 &      &      &      & -3.53 &    \\
\hspace{1em}Ret.($\omega=0$)   &      &      & -42.59 &      &      &      & -42.56 &      &      &      & -0.03 &    \\
\hspace{1em}Recoil   &      &      & 0.09 &      &      &      & 0.09 &      &      &      & 0.00 &    \\
Breit($\omega>0$)   & -9.02 &      & -8.98 &      & -8.47 &      & -8.43 &      & -0.55 &      & -0.55 &    \\
VP   &      &  &      &      &      &      &      &      &      &      &      &    \\
\hspace{1em}VP11   &      & -71.41 & -71.15 & -72.04 &      & -71.50 & -71.24 & -72.15 &      & 0.09 & 0.09 & 0.11\\
\hspace{1em}VP11+21   & -71.97 &      & -71.70 &      & -72.07 &      & -71.80 &      & 0.10 &      & 0.10 &    \\
\hspace{1em}VP11+21+13   &      & -69.11 & -69.09 &      &      & -69.20 & -69.18 &      &      & 0.09 & 0.09 &    \\
SE   &      &      &      &      &      &      &      &      &      &      &      &   \\
\hspace{1em}Welt.   & 376.72 &      & 376.48 & 375.28 & 376.43 &      & 376.16 & 375.06 & 0.29 &      & 0.31 & 0.22\\
\hspace{1em}dens.   & 382.41 &      &      &      & 382.08 &      &      &      & 0.32 &      &      &    \\
\hspace{1em}mod.   &      & 376.96 &      &      &      & 376.65 &      &      &      & 0.31 &      &    \\
QED h.o.   &      &      & -1.19 &      &      &      & -1.18 &      &      &      & 0.00 &    \\
Total(Welt.)   & -345702.86 &      & -345699.59 & -345674.67 & -346051.96 &      & -346048.73 & -346024.07 & 349.11 &      & 349.14 & 349.39\\
Total(dens.)   & -345697.17 &      &      &      & -346046.31 &      &      &      & 349.14 &      &      &    \\
Total(mod.)   &      & -345699.75 &      &      &      & -346048.87 &      &      &      & 349.13 &      &    \\
\bottomrule
\end{tabular*}
\end{sidewaystable}

\section{Theoretical background}

\subsection{MCDHF methods}

The methodology of MCDHF calculations performed in the present studies is similar to the one published earlier, in several papers (see, e.g., \cite{Grant1988,Grant2007,Grant2010,Grant1970,Gorceix1987,Indelicato1987}). The effective Hamiltonian for an $N$-electron system is expressed by
\begin{equation}
\hat H = \sum_{i=1}^{N} \hat h_{D}(i) + \sum_{j>i=1}^{N} V_{ij} 
\end{equation}
where $\hat h_{D}(i)$ is the Dirac one-particle operator for $i$-th electron and the terms $\hat V_{ij}$ account for the effective electron-electron interactions. 

An atomic state function (ASF) with the total angular momentum $J$, its $z$-projection $M$, and parity $p$ is assumed in the form
\begin{equation}
\Psi_{s} (JM^{p} ) = \sum_{m} c_{m} (s) \Phi ( \gamma_{m} JM^{p} )
\end{equation}
where $\Phi ( \gamma_{m} JM^{p} )$ are configuration state functions (CSF), $c_{m} (s)$ are the configuration mixing coefficients for 
state $s$, $\gamma_{m}$ represents all information  required  to uniquely define a certain CSF.
The CSF is a Slater determinant of Dirac 4-component bispinors:
\begin{equation}
\Phi(\gamma_m JM^p) = 
\sum_i d_i 
\begin{vmatrix}
\psi_1(1) & \cdots & \psi_1(N)\\
\vdots & \ddots & \vdots \\
\psi_N(1) & \cdots & \psi_N(N)
\end{vmatrix}
\end{equation}
where the $\psi_i$ is the one-electron wavefunctions and the $d_i$ coefficients are determined by requiring that the CSF is an eigenstate of $\hat{J^2}$ and $\hat{J_z}$. The one-electron wavefunction is defined as
\begin{equation}
\psi=
\frac{1}{r}
\begin{pmatrix}
P_{n,\kappa}(r)\cdot \Omega_{\kappa, j}^{m_j}(\theta,\phi)\\[0.8ex]
i Q_{n,\kappa}(r)\cdot \Omega_{-\kappa, j}^{m_j}(\theta,\phi)
\end{pmatrix}
\end{equation}
where $\Omega_{\kappa, j}^{m_j}(\theta,\phi)$ is a angular 2-component spinor and $P_{n,\kappa}(r)$ and $Q_{n,\kappa}(r)$ are large and small radial part of the wavefunction, respectively.

On the whole, the multiconfiguration DHS method is similar to the MCDF method, referring to effective Hamiltonian and multiconfigurational ASF. The main difference between the Dirac--Hartree--Fock method and the Dirac--Hartree--Fock--Slater method is that, in the (Dirac--)Hartree--Fock--Slater approach the nonlocal (Dirac--)Hartree--Fock exchange potential is approximated by a local potential. The FAC code uses an improved form of the local exchange potential (see \cite{Gu2008} for details). 

\subsection{Breit interaction}

The electron-electron interaction term is a sum of the Coulomb interaction $\hat V_{ij}^C$ operator and the transverse Breit $\hat V_{ij}^B$ operator \cite{Breit1929,Breit1930,Breit1932}:
\begin{equation}
\hat V_{ij} = \hat V_{ij}^C + \hat V_{ij}^B 
\label{eq:el-int}
\end{equation}
where the Coulomb interaction operator is  $\hat V_{ij}^C = 1/r_{ij}$, and the Breit operator in the Coulomb gauge is
\begin{equation}
{\hat V}_{ij}^B  =  - \bm{\alpha}_i \cdot \bm{\alpha}_j \frac{e^{i \omega_{ij} r_{ij}}}{r_{ij}} - ( \bm{\alpha}_i \cdot \bm{\nabla}_i ) (\bm{\alpha}_j \cdot \bm{\nabla}_j) \frac{e^{i \omega_{ij} r_{ij}} - 1}{\omega^2_{ij}r_{ij}} 
\label{eq:V_B}
\end{equation}
where $\omega_{ij} = (\varepsilon_i-\varepsilon_j)/c$ is the frequency
of one virtual photon exchanged and $\varepsilon_i$ and $\varepsilon_j$ are orbital energies of interacting electrons. 

The unretarded (instantaneous) parts are obtained making $\omega_{ij}$ $\rightarrow 0$. Then the Breit terms are given as
\begin{equation}
\label{eq:breit}
\begin{split}
\hat V_{ij}^B =& -\frac{\bm{\alpha}_i \cdot \bm{\alpha}_j}{2 r_{ij}} - \frac{(\bm{\alpha}_i \cdot r_{ij})(\bm{\alpha}_j \cdot r_{ij})}{2 r_{ij}^3} \\
&= \underbrace{-\frac{\bm{\alpha}_i \cdot \bm{\alpha}_j}{r_{ij}}}_{V_{mag}} + \underbrace{\left( \frac{\bm{\alpha}_i \cdot \bm{\alpha}_j}{2 r_{ij}} - \frac{(\bm{\alpha}_i \cdot r_{ij})(\bm{\alpha}_j \cdot r_{ij})}{2 r_{ij}^3} \right)}_{V_{ret}}
\end{split}
\end{equation}
where $V_{mag}$ is called magnetic (Gaunt) \cite{Gaunt1929} part and $V_{ret}$ is called retardation part. 

The zero-frequency approximation to the full transverse Breit interaction, i.e. Eq.~\eqref{eq:breit}, is well suited for most computations of many-electron atomic systems since the explicit frequency-dependent form, because of remedying the lack of covariance of Dirac-Coulomb-Breit Hamiltonian and the differences of state energy by using frequency-independent and frequency-dependent Breit operator are usually small \cite{Gorceix1987,Grant1976,Lindroth1989}.  
The Breit interaction can be included in two general ways: in the self-consistent field process, such as in \mcdfgme code \cite{Indelicato,Desclaux1975,Desclaux1984,Gorceix1987}, or in perturbational approach, such as in \textsc{Grasp}/\grasp codes \cite{Dyall1989,Jonsson2007}.

\subsection{QED corrections}

The bound-state vacuum polarization (VP) contribution is related to the creation and annihilation of virtual electron-positron pairs in the field of the nucleus. It is a correction to the photon propagator. 
The first term of order $\alpha(Z\alpha)$ can be calculated as the expectation value of the Uehling potential. 
The Uehling potential in the case of finite nuclear size and spherical symmetric nuclear charge distribution $\rho(\vec r)$ can be 
expressed as \cite{Fullerton1976}:
\begin{equation}
\begin{array}{>{\displaystyle}c}
U(\vec r) = -\frac{2}{3}\frac{Z\alpha^2 \hbar^2}{mr} \int_0^\infty d^3r'\, r' \rho(r') \\[1ex]
\times \left[ K_0 \left(\frac{2mc}{\hbar}|r-r'|\right) - K_0 \left(\frac{2mc}{\hbar}|r+r'|\right) \right]
\end{array}
\end{equation}
where the function $K_0(x)$ is defined as:
\begin{equation}
K_0(x) = \int_1^\infty dt\, e^{-xt} \left( \frac{1}{t^3}+\frac{1}{2t^5} \right) \sqrt{t^2-1}
\end{equation}
The higher-order terms have been given by K{\"a}ll{\'e}n and Sabry \cite{Kallen1955} for order~$\alpha^2(Z\alpha)$ and by Wichmann and Kroll \cite{Wichmann1956,Blomqvist1972} for order~$\alpha(Z\alpha)^3$.

Self-energy (SE) contribution arises from the interaction of the electron with its own radiation field. It is a correction to the electron propagator. 
For one-electron systems the most important (one-loop) self-energy term has been calculated exactly by Mohr \cite{Mohr1975,Mohr1982,Mohr1992} and expressed as:
\begin{equation}
\Delta E_{n\kappa} = \frac{\alpha}{\pi} \frac{(Z\alpha)^4}{n^3} F_{n\kappa}(Z\alpha) \;m_ec^2
\end{equation}
where $F_{n\kappa}(Z\alpha)$ is a slowly varying function of $Z\alpha$. 
For many-electron atomic systems the self-energy correction to the energy is changed by the electron screening. There are three general ways to estimate self-energy screening for atoms. The major differences between these approaches are for results of SE correction to the energy of $s$ subshells. 

In the 'Welton picture' approach \cite{Welton1948,Indelicato1987,Indelicato1990} the self-energy correction for $s$-type Dirac-Fock orbitals is scaled from exact hydrogenic results from the following relation:
\begin{equation}
\left(\Delta E_{ns}\right)_{DF} = \frac{\langle ns | \nabla^2 V_{nucl}(r) | ns \rangle_{DF}}{\langle ns | \nabla^2 V_{nucl}(r) | ns \rangle_{Hyd}} \left(\Delta E_{ns}\right)_{Hyd}
\label{SEscr-welt}
\end{equation}
where $V_{nucl}(r)$ is a nuclear potential. This approach is implemented in \mcdfgme code. Lowe \textit{et al.} \cite{Lowe2013} created extension of \grasp package, that implements Welton picture approach to estimate SE screening into \grasp suite. 
The \grasp code natively approximates the screening coefficient by taking the ratio of the Dirac-Fock wavefunction density in a small region around the nucleus ($r\le r'$, $r'=0.0219 a_0$, $a_0$ -- Bohr's radius) to the equivalent density for a hydrogenic orbital, i.e. \cite{Lowe2013}
\begin{equation}
\left(\Delta E_{nl}\right)_{DF} = \frac{\langle nl_{r\le r'} | nl_{r\le r'} \rangle_{DF}}{\langle nl_{r\le r'} | nl_{r\le r'} \rangle_{Hyd}} \left(\Delta E_{nl}\right)_{Hyd}
\label{SEscr-dens}
\end{equation}
This approach is called 'density approach' further in the manuscript. 
Last years some modern approaches for the estimation of hydrogenic SE data to many-electron atoms have been presented, such as the model Lamb-shift operator \cite{Flambaum2005,Thierfelder2010,Shabaev2015} and the spectral representation (projection operator) of the Lamb shift \cite{Dyall2013}. 
Recently Shabaev \textit{et al.} \cite{Shabaev2015,Shabaev2018} published \qedmod, a program for calculating the model Lamb-shift operator basing on numerical radial wavefunctions. In this paper the \grasp wavefunctions are used as a \qedmod input.

\begin{sidewaystable*}
\setlength{\tabcolsep}{4pt}
\caption{\label{tab:ka12}Various theoretical contributions to the energy of $1s^{-1}$, $2p_{1/2}^{-1}$, and $2p_{3/2}^{-1}$ hole states of \Wviii (eV) and energy of $K\alpha_{1,2}$ transitions.}
\small
\begin{tabular*}{\linewidth}{@{\extracolsep{\fill}}l *{12}{>{\small\arraybackslash}r}}
\toprule
& \multicolumn{4}{c}{$1s^{-1}$} & \multicolumn{4}{c}{$2p_{1/2}^{-1}$} & \multicolumn{4}{c}{$2p_{3/2}^{-1}$} \\
\cmidrule{2-5}\cmidrule{6-9}\cmidrule{10-13}
 &   \multicolumn{1}{c}{\grasp}  &  \multicolumn{1}{c}{\grasp}  &   \multicolumn{1}{c}{\mcdfgme}  &   \multicolumn{1}{c}{\fac}  &   \multicolumn{1}{c}{\grasp}  &  \multicolumn{1}{c}{\grasp}  &   \multicolumn{1}{c}{\mcdfgme}  &   \multicolumn{1}{c}{\fac}  &   \multicolumn{1}{c}{\grasp}  &  \multicolumn{1}{c}{\grasp}  &   \multicolumn{1}{c}{\mcdfgme}  &   \multicolumn{1}{c}{\fac} \\
Contribution  &    &  \multicolumn{1}{c}{+\qedmod}  &    &    &    &  \multicolumn{1}{c}{+\qedmod}  &    &    &    &  \multicolumn{1}{c}{+\qedmod}  &    &   \\
\midrule
Dirac-Fock  & -369478.91 &  & -369476.90 & -369453.39 & -427763.39 &  & -427760.56 & -427729.68 & -429115.93 &  & -429112.94 & -429082.02\\
Breit($\omega=0$)+Rec.  & 234.57 &  & 234.29 & 234.08 & 425.12 &  & 424.64 & 425.46 & 440.54 &  & 439.40 & 441.22\\
\hspace{1em}Mag.  &    &  & 270.66 &    &    &  & 474.72 &    &    &  & 489.49 &   \\
\hspace{1em}Ret.($\omega=0$)  &    &  & -36.43 &    &    &  & -50.18 &    &    &  & -50.20 &   \\
\hspace{1em}Recoil  &    &  & 0.06 &    &    &  & 0.10 &    &    &  & 0.10 &   \\
Breit($\omega>0$)  & -4.95 &  & -5.00 &    & -9.05 &  & -9.08 &    & -7.24 &  & -7.26 &   \\
VP  &    &  &    &    &    &  &    &    &    &  &    &   \\
\hspace{1em}VP11  &    & -43.31 & -42.27 & -42.59 &    & -72.34 & -71.11 & -71.99 &    & -72.69 & -71.42 & -72.36\\
\hspace{1em}VP11+21  & -42.61 &  & -42.56 &    & -71.91 &  & -71.66 &    & -72.26 &  & -71.98 &   \\
\hspace{1em}VP11+21+13  &    & -40.90 & -41.01 &    &    & -69.05 & -69.06 &    &    & -69.38 & -69.35 &   \\
SE  &    &  &    &    &    &  &    &    &    &  &    &   \\
\hspace{1em}Welt.  & 232.51 &  & 231.23 & 228.50 & 379.95 &  & 377.50 & 375.58 & 378.85 &  & 376.07 & 374.49\\
\hspace{1em}dens.  & 232.56 &  &    &    & 382.62 &  &    &    & 381.52 &  &    &   \\
\hspace{1em}mod. &  & 230.35 &  &  &  & 377.68 &  &  &  & 376.39 &  & \\
QED h.o.  &    &  & -1.05 &    &    &  & -1.26 &    &    &  & -1.25 &   \\
Total(Welt.)  & -369059.38 &  & -369058.44 & -369033.40 & -427039.28 &  & -427037.80 & -427000.63 & -428376.04 &  & -428375.32 & -428338.66\\
Total(dens.)  & -369059.33 &  &    &    & -427036.62 &  &    &    & -428373.37 &  &    &   \\
Total(mod.) &  & -369059.84 &  &  &  & -427038.69 &  &  &  & -428375.63 &  & \\
\midrule
& \multicolumn{4}{c}{$K\alpha_1$} & \multicolumn{4}{c}{$K\alpha_2$} &&&&\\
\cmidrule{2-5}\cmidrule{6-9}
 &   \multicolumn{1}{c}{\grasp}  &  \multicolumn{1}{c}{\grasp}  &   \multicolumn{1}{c}{\mcdfgme}  &   \multicolumn{1}{c}{\fac}  &   \multicolumn{1}{c}{\grasp}  &  \multicolumn{1}{c}{\grasp}  &   \multicolumn{1}{c}{\mcdfgme}  &   \multicolumn{1}{c}{\fac} \\
Contribution  &    &  \multicolumn{1}{c}{+\qedmod}  &    &    &    &  \multicolumn{1}{c}{+\qedmod}  &    &  &&&& \\
\midrule
Dirac-Fock  & 59637.02 &  & 59636.04 & 59628.63 & 58284.48 &  & 58283.65 & 58276.29 &  &  &  & \\
Breit($\omega=0$)+Rec.  & -205.97 &  & -205.11 & -207.14 & -190.54 &  & -190.35 & -191.38 &  &  &  & \\
\hspace{1em}Mag.  &    &  & -218.84 &    &    &  & -204.07 &    &  &  &  & \\
\hspace{1em}Ret.($\omega=0$)  &    &  & 13.77 &    &    &  & 13.76 &    &  &  &  & \\
\hspace{1em}Recoil  &    &  & -0.04 &    &    &  & -0.04 &    &  &  &  & \\
Breit($\omega>0$)  & 2.29 &  & 2.26 &    & 4.10 &  & 4.08 &    &  &  &  & \\
VP  &    &  &    &    &    &  &    &    &  &  &  & \\
\hspace{1em}VP11  &    & 29.42 & 29.19 & 29.77 &    & 29.07 & 28.88 & 29.40 &  &  &  & \\
\hspace{1em}VP11+21  & 29.66 &  & 29.42 &    & 29.30 &  & 29.11 &    &  &  &  & \\
\hspace{1em}VP11+21+13  &    & 28.48 & 28.35 &    &    & 28.14 & 28.05 &    &  &  &  & \\
SE  &    &  &    &    &    &  &    &    &  &  &  & \\
\hspace{1em}Welt.  & -146.34 &  & -144.85 & -145.99 & -147.44 &  & -146.27 & -147.08 &  &  &  & \\
\hspace{1em}dens.  & -148.96 &  &    &    & -150.06 &  &    &    &  &  &  & \\
\hspace{1em}mod. &  & -146.04 &  &  &  & -147.33 &  &  &  &  &  & \\
QED h.o.  &    &  & 0.20 &    &    &  & 0.21 &    &  &  &  & \\
Total(Welt.)  & 59316.66 &  & 59316.88 & 59305.26 & 57979.90 &  & 57979.36 & 57967.23 &  &  &  & \\
Total(dens.)  & 59314.04 &  &    &    & 57977.28 &  &    &    &  &  &  & \\
Total(mod.) &  & 59315.79 &  &  &  & 57978.85 &  &  &  &  &  & \\
\bottomrule
\end{tabular*}
\end{sidewaystable*}

\section{Results and discussion}

\subsection{Breit and QED contributions}

In present work the \fac, \grasp, and \mcdfgme codes are compared in three case studies of radiative transitions occurring in tungsten ions. 
The first case study is focused on radiative transitions among outer orbitals in highly ionised tungsten. The study of characteristic x-ray radiation emitted by highly ionized W atoms is of great importance for both theoretical and applied atomic physics, including fusion applications \cite{Gillaspy2010,Podpaly2009,Ralchenko2006}. 
In the recent works of Rzadkiewicz \textit{et al.} \cite{Rzadkiewicz2018} and Kozioł and Rzadkiewicz \cite{Kozio2018}, the energy levels of the ground and excited states of \WCu and \WNi ions and the wavelengths and transition probabilities of the $4d \to 3p$ transitions were calculated using the MCDHF Configuration Interaction (CI) method, but no deeper analysis of the Breit and QED contributions to the energy levels was performed. 
Hence, the Ni1 and Ni2 transitions in Ni-like (\WNi) tungsten were selected as the first case study. The initial levels of these transitions are $\ket{[Mg]3p_{1/2}^2 3p_{3/2}^3 3d^{10} 4d_{3/2}^1}_{J=1}$ ($^3P_{1}$) and $\ket{[Mg]3p_{1/2}^2 3p_{3/2}^3 3d^{10} 4d_{5/2}^1}_{J=1}$ ($^1P_{1}$) and the final level is $\ket{[Mg]3p^63d^{10}}_{J=0}$ ($^1S_{0}$). The Ni1 and Ni2 transitions are ${}^1P_1 \to {}^1S_0$ and ${}^3P_1 \to {}^1S_0$, respectively. 

The second case study concerns the $3p_{3/2}{-}3p_{1/2}$ fine splitting in Cl-like (\WCl) tungsten. 
Recently, a Cl-like isoelectronic sequence was proposed as an electronic configuration that could be used to accurately test current methods of computing the Breit and QED effects \cite{Bilal2018}. 
The $3p_{3/2}{-}3p_{1/2}$ fine splitting in Cl-like tungsten has recently been investigated both experimentally \cite{Lennartsson2013} and theoretically \cite{Quinet2011,Aggarwal2014,Singh2016}. 

The third case study is focused on core radiative transitions in stripped tungsten. 
The energy shifts of the $K\alpha_{1,2}$, $K\beta_{1,3}$ and $K\beta_2$ lines of stripped high-$Z$ atoms have been suggested as being potentially relevant to diagnostics of high-energy-density laser-produced plasmas \cite{Rzadkiewicz2013}. 
Hence, the $K\alpha_{1}$ ($1s^{-1}\to2p_{3/2}^{-1}$) and $K\alpha_{2}$ ($1s^{-1}\to2p_{1/2}^{-1}$) transitions in W$^{7+}$ were selected as the third case study. 
The W$^{7+}$ ion was selected because it has simple closed-shell valence electronic configurations and, as a result, there are only two transitions between initial and final states to be studied. The electronic configurations for the $1s^{-1}$, $2p_{1/2}^{-1}$, and $2p_{3/2}^{-1}$ hole states are then $1s_{1/2}^1 2s^2 2p^6 M^{18} N^{32} 5s^2 5p^6$, $1s^2 2s^2 2p_{1/2}^1 2p_{3/2}^4 M^{18} N^{32} 5s^2 5p^6$, and $1s^2 2s^2 2p_{1/2}^2 2p_{3/2}^3 M^{18} N^{32} 5s^2 5p^6$, respectively.

Table~\ref{tab:ni12} presents various theoretical contributions to the energies of the $[Mg]3p^{5}3d^{10}4d^{1}(J=1)$ and $[Mg]3p^{6}3d^{10}(J=0)$ states of the Ni-like W ion and the energies of the Ni1 and Ni2 transitions. 
Table~\ref{tab:cl3p} presents various theoretical contributions to the energy of the $3p_{3/2}-3p_{1/2}$ fine splitting in a Cl-like tungsten ion. 
Table~\ref{tab:ka12} presents various theoretical contributions to the energies of $1s^{-1}$, $2p_{1/2}^{-1}$, and $2p_{3/2}^{-1}$ hole states of \Wviii and the $K\alpha_{1}$ and $K\alpha_{2}$ transitions. 
Relativistic and radiative effects are treated slightly differently in the different codes. 
The term 'Dirac-Fock' in the tables indicates that the energy of the state was obtained using the self-consistent multiconfigurational Dirac-Hartree-Fock (\grasp and \mcdfgme codes) or the Dirac-Hartree-Fock-Slater (\fac code) procedure, without higher order corrections. 
'Breit($\omega=0$)+Rec.' refers to the Breit correction in the zero-frequency approximation (see, e.g., \cite{Kozio2018b} and references therein for details) plus the recoil correction. Using the \mcdfgme code, it is also possible to print the particular contributions included in this correction; they are: 'Mag.' -- magnetic (Gaunt) part of the Breit interaction, 'Ret.($\omega=0$)' -- retardation (gauge) part of the Breit interaction in the zero-frequency approximation, and 'Recoil' -- recoil correction (including relativistic recoil). 'Breit($\omega>0$)' indicates the retardation term of the Breit interaction beyond the zero-frequency approximation. This aspect is not included in the \fac calculations, which is common in many atomic computational codes. 
Each of studied codes has a different treatment of VP correction. The \mcdfgme code takes into account three VP potentials: V11 (Uehling potential), V21 (K{\"a}ll{\'e}n and Sabry potential), and V13 (Wichmann and Kroll potential). In the \fac code only V11 is included while the \grasp code takes into account the sum of V11 and V21. 
The different methods for including the SE correction in many-electron atomic systems that have been used are the density approach ('dens.' in Tables), the Welton picture approach ('Welt.' in Tables), and the model Lamb-shift operator ('mod.' in Tables). 

As seen from Tables~\ref{tab:ni12}, \ref{tab:cl3p}, and \ref{tab:ka12}, the absolute level energies calculated by the \fac code differ significantly from the energies calculated by the \grasp and \mcdfgme codes, by about 20--30~eV. However, the energies of the radiative transitions differ much less, about 1~eV in the case of the Ni1 and Ni2 lines and below 1~eV for the $3p_{3/2}-3p_{1/2}$ fine splitting in \WCl. For the $K\alpha_{1}$ and $K\alpha_{2}$ transitions in \Wviii, the \fac calculated numbers are smaller by about 10~eV than those obtained by the MCDHF codes. This difference is too large to estimate properly the outer-shell ionization level from $K$-shell x-ray lines shift \cite{Rzadkiewicz2013}. 
It can be concluded that the \fac code is sufficiently accurate in cases where radiative transitions are linked to an electron jump within the valence shells (if high accuracy is not required), but is less accurate for transitions that are linked to inner-shell hole states. 

Comparing the Breit contributions obtained from the \grasp code (where the Breit term is treated perturbatively) and from the \mcdfgme code (where Breit term is included in a variational SCF process) allows the so-called ''variational effect'' to be estimated. The magnitude of this effect is about 1~eV (0.1--0.3\%) in the cases studied here. However, it has been found that  the variational effect is significantly reduced when active space is expanding \cite{Si2018,Chantler2014}. The frequency-dependent Breit term is about 2\% of the frequency-independent one (having the opposite sign). 

As mentioned above, three different approximations to estimate the SE corrections have been used: the Welton picture, the density approach, and the model Lamb-shift operator. For \grasp calculations, it is possible to compare these models by using these same wavefunctions. 
In the case of \WNi, the 'density' approach gives SE contributions to the energy levels that are significantly larger, by about 5~eV, than those of the other two approaches. However, this difference vanishes for the Ni1 and Ni2 transition energies. The case of the \WCl ion is similar. 
For the $1s^{-1}$, $2p_{1/2}^{-1}$, and $2p_{3/2}^{-1}$ hole states of \Wviii the 'model operator' method gives SE contributions to the energy levels that are significantly smaller than those of the 'Welton picture', which in turn are smaller than those of the 'density' approach.

\subsection{Relative role of correlation contribution and Breit and QED contributions}

It is interesting to compare the electron correlation contributions to the Breit and QED contributions in selected cases. 
For the Ni1 and Ni2 lines, the correlation contribution was studied extensively by Rzadkiewicz \textit{et al.} \cite{Rzadkiewicz2018} and Kozioł and Rzadkiewicz \cite{Kozio2018} using a MCDHF Configuration Interaction (CI) calculation. They pointed out that electron correlation effect ranges from -1.87~eV to -2.87~eV for the Ni1 line energy and from -1.05~eV to -2.45~eV for the Ni2 line energy, depending on the CI model used. The correlation effect is then larger by an order of magnitude than the frequency-dependent Breit term (omitted in the calculations in \cite{Rzadkiewicz2018,Kozio2018} due to the inclusion of virtual orbitals within the CI procedure) and larger by more than an order of magnitude than the differences arising from the use of different QED models.  

\begin{table}[!htb]
\caption{\label{tab:cl3pci}The $3p_{3/2}-3p_{1/2}$ fine splitting calculated for various CI active spaces.}
\centering
\begin{tabular*}{\linewidth}{@{\extracolsep{\fill}}lll}
\toprule
Active space & Energy (eV) & Wavelength (\AA)\\
\midrule
AS0 & 349.11 & 35.515 \\
AS1 & 347.26 & 35.703 \\
AS2 & 347.84 & 35.644 \\
AS3 & 347.62 & 35.666 \\
AS4 & 347.50 & 35.678 \\
AS5 & 347.49(1) & 35.680(2) \\
\midrule
\multicolumn{2}{l}{Experiment:} & \\
\multicolumn{2}{l}{\hspace{1em} Lennartsson \textit{et al.} \cite{Lennartsson2013}} & 35.668(4)\\
\multicolumn{2}{l}{Other theory:} & \\
\multicolumn{2}{l}{\hspace{1em} Quinet \cite{Quinet2011}} & 35.633\\
\multicolumn{2}{l}{\hspace{1em} Singh and Puri \cite{Singh2016}} & 35.632\\
\multicolumn{2}{l}{\hspace{1em} Aggarwal and Keenan \cite{Aggarwal2014}} & 35.686\\
\multicolumn{2}{l}{\hspace{1em} Bilal \textit{et al.} \cite{Bilal2018}} & 35.765\\

\bottomrule
\end{tabular*}
\end{table}

In the case of the $3p_{3/2}-3p_{1/2}$ fine splitting in \WCl the MCDHF-CI calculations were performed with the \grasp code (see e.g. \cite{Rzadkiewicz2018,Kozio2018} for details). 
The $1s$, $2s$, and $2p$ subshells are inactive orbitals. All single (S) and double (D) substitutions from the $3s$ and $3p$ orbitals to the active spaces (AS) of virtual orbitals are allowed. 
The virtual orbital sets used were: AS1 = \{3d,4s,4p,4d,4f\}, AS2 = AS1 + \{5s,5p,5d,5f,5g\}, AS3 = AS2 + \{6s,6p,6d,6f,6g\}, AS4 = AS3 + \{7s,7p,7d,7f,7g\}, and AS5 = AS4 + \{8s,8p,8d,8f,8g\}. 
Table~\ref{tab:cl3pci} collects the results of the $3p_{3/2}-3p_{1/2}$ fine splitting calculated for various CI active spaces. The AS0 value is a number related to \grasp calculations with the 'Welton picture' approach for estimating the SE contribution. 
For the final active space, AS5, the theoretical uncertainties are presented. They are related to convergence with the size of a basis set and estimated as an absolute value of difference between energy(wavelengths) calculated for AS4 and AS5 stages. 
The wavelengths for the AS3-AS5 approaches agree well with the experimental values from the work of Lennartsson \textit{et al.} \cite{Lennartsson2013}. 
The correlation effect is about 1.62~eV, which is about three times larger than the frequency-dependent Breit term and the total QED effect.

\begin{table}[!htb]
\setlength{\tabcolsep}{4pt}
\caption{\label{tab:ka12ci}The $K\alpha_{1}$ and $K\alpha_{2}$ transitions energy in \Wviii calculated for various CI active spaces. Theoretical uncertainties for 'AS3 + Auger shift' are related to convergence with the size of a basis set and to error of interpolation Auger shift corrections.}
\centering
\begin{tabular*}{\linewidth}{@{\extracolsep{\fill}}lll}
\toprule
Active space & \multicolumn{2}{c}{Energy (eV)}\\
\cmidrule{2-3}
& $K\alpha_{1}$ & $K\alpha_{2}$ \\
\midrule
AS0 & 59316.66 & 57979.90 \\
AS1 & 59315.82 & 57979.06 \\
AS2 & 59316.43 & 57979.60 \\
AS3 & 59316.63 & 57979.80 \\
AS3 + Auger shift & 59318.77(30) & 57983.13(30) \\
\midrule
Experiment (neutral W): &  &  \\
\hspace{1em} Bearden \cite{Bearden1967} & 59318.24 & 57981.7 \\
\hspace{1em} Deslattes \textit{et al.} \cite{Deslattes2003} & 59318.847(50) & 57981.77(14) \\
Other theory (neutral W): &  &  \\
\hspace{1em} Deslattes \textit{et al.} \cite{Deslattes2003} & 59318.8(17) & 57981.9(19) \\
\bottomrule
\end{tabular*}
\end{table}

For the $K\alpha_{1}$ and $K\alpha_{2}$ transition energies in \Wviii, MCDHF-CI calculations were performed to check correlation effects.  The active space of occupied orbitals contains orbitals involved in radiative transitions ($1s$ and $2p$) and two the most outer subshells: $4f$ and $5p$. All other occupied subshells are inactive core. All SD substitutions from the active space of occupied orbitals to the active spaces of virtual orbitals are allowed. The virtual orbital active spaces used were: AS1 = \{5d,5f,5g\}, AS2 = AS1 + \{6s,6p,6d,6f,6g\}, and AS3 = AS2 + \{7s,7p,7d,7f,7g\}. 
The results of the $K\alpha_{1}$ and $K\alpha_{2}$ transition energies in \Wviii, calculated with various CI active spaces, are shown in Table~\ref{tab:ka12ci}. It is evident that the correlation effects are small in this case. Similar order of magnitude for correlation effects were found for the $K\alpha_{1,2}$ lines of Al and Si~\cite{Kozio2014b} and of Kr, Pb, U, Pu, Fm \cite{Indelicato1992,Lindroth1993}. 
Because inner-hole states are autoionising, the level energy shift due to the coupling between the two hole states and one excited electron, called Auger shift, must be included \cite{Deslattes2003}. 
The Auger shift contribution to the $K\alpha_{1}$ and $K\alpha_{2}$ transition energies can be approximated through interpolation from the numbers given by Indelicato, Lindroth \textit{et al.} \cite{Indelicato1992,Mooney1992,Lindroth1993}. 
The interpolated values for the Auger shift contribution are 2.14~eV and 3.33~eV for the $K\alpha_{1}$ and $K\alpha_{2}$ lines, respectively. 
Adding the Auger shift contributions to the MCDHF-CI values reduces the substantially discrepancy between theory and experiment for $K\alpha_{1}$ line, however overestimate value for $K\alpha_{2}$ line.

\section{Conclusions}

The interpretation of atomic observations by theory and the testing of computational predictions by experiment are interactive processes. 
In this paper the \fac, \grasp, and \mcdfgme codes have been compared in selected case studies involving radiative transitions occurring in the tungsten ions \Wviii, \WCl, and \WNi. 
Transitions involving electron jumps between outer or inner orbitals are both considered. 
Various approaches to including the Breit interaction term and QED corrections in atomic calculations have been examined and their contributions compared to those of electron correlations. 
In the case when transitions involve electron jumps between outer shells (the first and second cases in the present work) the frequency-dependent Breit contribution to transition energy is smaller few times than the electron correlation contribution. Then, ommiting the frequency-dependent Breit term is not a big mistake. In the case when transitions involve electron jumps between inner and outer shells (the third case in the present work) the frequency-dependent Breit contribution dominates over the electron correlation contribution. In this case also the differences between QED models may be bigger than correlation contribution. 
The presented data may be used to estimate theoretical uncertainties relevant to interpretation of high-resolution spectroscopic data.

\section*{Acknowledgments}

The work was partly supported by the Polish Ministry of Science and Higher Education within the framework of the scientific financial resources in the years 2016--2019 allocated for the realization of the international co-financed project. This work has been carried out within the framework of the EUROfusion Consortium and has received funding from the Euratom Research and Training Programme 2014--2019 under Grant Agreement No. 633053. The views and opinions expressed herein do not necessarily reflect those of the European Commission.

%

\end{document}